\magnification=\magstep1
\baselineskip=20pt plus.1pt              
\pageno=1
\centerline      {\sl\bf IMPURITY SCATTERING IN A D-WAVE SUPERCONDUCTOR}
\def\esn{\epsilon} 
 
\def\vq{\vec q} 
\def\vk{\vec k} 
\def\vr{\vec r}
\def\br{{\bf r}}
\def\bs{{\bf s}}
\def\bR{\bf R}
\def\bS{\bf S}
\def\vS{\vec S}
\def\hb{\hfill\break}
\vskip 2pc
\centerline {Mi-Ae Park and  M. H. Lee} 
\centerline {Department of Physics, Seoul National University}
\centerline {Seoul 151-742, Korea}
\vskip 1pc
\centerline {Yong-Jihn Kim} 
\centerline {Department of Physics, Korea Advanced Institute of Science and Technology}
\centerline {Taejon 305-701, Korea}
\vskip 2pc
The influence of (non-magnetic and magnetic) impurities on the transition 
temperature of a d-wave superconductor is studied anew within the
framework of BCS theory.
Pairing interaction decreases linearly with the impurity concentration. 
Accordingly $T_{c}$ suppression is proportional to the 
(potential or exchange) scattering rate, $1/\tau$, due to impurities. 
The initial slope versus $1/\tau$ is found to depend
on the superconductor contrary to Abrikosov-Gor'kov type theory.
Near the critical impurity concentration $T_{c}$ drops abruptly to zero.
Because the potential scattering rate is generally much larger than 
the exchange scattering rate, magnetic impurities will also act as
non-magnetic impurities as far as the $T_{c}$ decrease is concerned.
The implication for the impurity doping effect in high $T_{c}$ superconductors
is also discussed.
 
\vskip 1.0in\hb
\vskip 2pc

\vfill\eject 

\centerline{\bf 1. Introduction}
\vskip 1pc
Recently there has been much attention on the d-wave pairing state
in connection with high-temperature superconductors.$^{1, 2, 3}$
The impurity effect on the d-wave pairing is particularly interesting 
because it may give a clue to the symmetry of the superconducting state
of the high $T_{c}$ cuprates. 
There are already many theoretical works on this issue.$^{4-12}$
For examples, the impurity effects on the penetration depth,$^{4}$
the superfluid density,$^{4}$ the transition temperature,$^{5}$ 
the quasiparticle states,$^{6}$ 
the density of states,$^{7,10-12}$  
the nuclear spin relaxation rate,$^{7}$  
the spin susceptibility,$^{8}$ and
the infrared conductivity,$^{9}$  have been investigated.

The above theoretical works were essentially based on
the Abrikosov-Gorkov's (AG) pair breaking theory$^{13}$ for the s-wave 
superconductors.  However, Kim and Overhauser$^{14}$ proposed 
a new theory with different predictions: (i) 
The initial slope of $T_{c}$ decrease depends on the superconductor and 
is not the universal constant suggested by AG. (ii) 
The $T_{c}$ reduction by exchange scattering is partially suppressed 
by potential scattering when the mean free path is smaller than the 
coherence length, which has been confirmed in several experiments.$^{15-17}$
Kim$^{18}$ also showed that the failure of AG theory comes from the intrinsic
pairing problem in Gor'kov's formalism, which 
can be cured by incorporating the pairing constraint.
The purpose of this paper is to reconsider the $T_{c}$ suppression
due to (non-magnetic and magnetic) impurities in a d-wave superconductor 
within the framework of BCS theory.  

It is shown that the pairing interaction decreases linearly with the
(non-magnetic or magnetic) impurity concentration. The ratio of the average
correlation length, $\xi_{o} = 0.18{v_{F}\over T_{c}}$, to
the mean free path, $\xi_{o}/ \ell$, determines the 
weakening of the pairing interaction. Consequently $T_{c}$ suppression
and the initial slope depend strongly on the superconductor   
contrary to Abrikosov-Gor'kov type theory.
For the magnetic impurities, because the cross section of the potential scattering  
is generally much larger than that of the exchange
scattering, the $T_{c}$ decrease is also determined
by the potential scattering of the magnetic impurities.$^{13,19}$ 

The interpretation of the high $T_{c}$ data$^{20,21}$ by the present study 
is not clear, though.
The observed $T_{c}$ decrease due to impurity scattering is much slower 
than the theory predicts.
The discrepancy seems to come from the neglect of the strong Coulomb
interaction effect in calculating the impurity response of the
superconducting state. In other words, the impurity
responses are different in Fermi liquids and strongly correlated systems.
Note that it was pointed out$^{22}$ the $T_{c}$ decrease caused by
impurity doping and ion-beam-induced damage in high $T_{c}$ superconductors 
is related with the proximity to a metal-insulator transition.

\vskip 1pc
\centerline{\bf 2. D-Wave Superconductor}
\vskip 1pc
For a d-wave superconductor, the pairing interaction $V_{\vk,\vk'}$
for the plane states is taken to be$^{23}$
$$V_{\vk,\vk'} = \int e^{i(\vk-\vk')\cdot \vr}V({\br})d^{3}r = - 5V_{2}
 \ {1\over 2}[3({\hat k}\cdot{\hat k'})^{2}-1],\eqno(1)$$
where $\hat k$ is the unit vector parallel to $\vk$.
Substituting Eq. (1) into the BCS gap equation, one finds
$$\Delta_{\vk} = 5V_{2}\sum_{\vk'}
 {1\over 2}[3({\hat k}\cdot{\hat k'})^{2}-1]
{\Delta_{\vk'} \over 2E_{\vk'}}
tanh{E_{\vk'}\over 2T}, \eqno(2)$$
where
$$E_{\vk'} = \sqrt{\esn_{\vk'}^{2} + |\Delta_{\vk'}|^{2}},\eqno(3)$$
and $\esn_{\vk}$ is the electron energy.
Among the possible solutions, we consider
$$\Delta_{\vk} = \Delta_{o}({\hat {k}_{x}}^{2} - {\hat {k}_{y}}^{2}).\eqno(4)$$
This solution has the same symmetry property as $d_{x^{2}-y^{2}}=\Delta_{o}
(cos k_{x}-cos k_{y})$ which is believed to describe the gap structure
of the cuprate high $T_{c}$ superconductors.
Upon substitution in Eq. (2), one obtains the $T_{c}$
equation
$$T_{c} = 1.13\esn_{c}e^{-1/ {N_{o}V_{2}}}, \eqno(5)$$
where $\esn_{c}$ is the cutoff energy and $N_{o}$ is the electronic 
density of states at the Fermi level.  

In the presence of impurities, the scattered states $\psi_{n}$ may be expanded
in terms of plane waves, such as$^{24}$
$$\psi_{n}=\sum_{\vec k}e^{i{\vec k}\cdot{\bf r}}<{\vec k}|n>.\eqno(6)$$
Now the pairing interaction $V_{nn'}$ between scattered basis pairs
$(\psi_{n}, \psi_{\bar n})$ and $(\psi_{n'}, \psi_{\bar n'})$ is given by
$$V_{nn'}=\int\int d{\bf r}_{1}d{\bf r}_{2}\psi_{n'}^{*}({\bf r}_{1})
\psi_{\bar n'}^{*}({\bf r}_{2})V(|{\bf r}_{1}-{\bf r}_{2}|)
\psi_{\bar n}({\bf r}_{2}) \psi_{n}({\bf r}_{1}).\eqno(7)$$
Here $\psi_{\bar n}$ denotes the time-reversed state of $\psi_{n}$.
From Eqs. (1), (6) and (7) we can calculate $V_{nn'}$.

\vskip 1pc
\centerline{\bf 3. Non-Magnetic Impurity Effect}
\vskip 1pc
The non-magnetic impurities will be examined first.
The scattering potential from the impurities is given
$$U({\br}) = \sum_{i}u\delta({\br} - {\bR}_{i}).\eqno(8)$$
$\{{\bR}_{i}\}$ is the impurity sites.
We consider the impurity effect using the 2-nd order perturbation theory. 
Then, the scattered state which carries the label $\vk\alpha$ is
$$\psi_{\vk\alpha} = N_{\vk}[e^{i\vk\cdot{\br}} + \sum_{i,\vq}{u\over 
\esn_{\vk} - \esn_{\vk+\vq}}e^{-i\vq\cdot\bR_{i}}e^{i(\vk+\vq)\cdot\br}]
\alpha,\eqno(9)$$
where $N_{\vk}$ is the normalizing factor.
The time-reversed degenerate partner of Eq. (9) is
$$\psi_{-\vk\beta} = N_{\vk}[e^{-i\vk\cdot{\br}} + \sum_{i,\vq}{u\over 
\esn_{\vk} - \esn_{\vk+\vq}}e^{i\vq\cdot\bR_{i}}e^{-i(\vk+\vq)\cdot\br}]
\beta.\eqno(10)$$

The matrix elements which cause Cooper pairing are
between scattered basis pairs $(\psi_{\vk\alpha},\psi_{-\vk\beta})$ 
and $(\psi_{\vk'\alpha},\psi_{-\vk'\beta})$.$^{14,24}$ 
Each basis pair is a $2\times 2$ Slater determinant.
The matrix element $V_{\vk,\vk'}$ between the two scattered
basis determinants is
$$V_{\vk,\vk'} = <{\cal D}_{\vk'}(\vr_{1},\vr_{2})\big|V(|\vr_{1}-\vr_{2}|)
\big| {\cal D}_{\vk}(\vr_{1},\vr_{2})>,\eqno(11)$$
where
$${\cal D}_{\vk}(\vr_{1},\vr_{2}) = 
{1\over \sqrt{2}} \left|\matrix{\psi_{\vk\alpha}({\br}_{1})&\psi_{\vk\alpha}
({\br}_{2})\cr
\psi_{-\vk\beta}({\br}_{1})&\psi_{-\vk\beta}({\br}_{2})\cr}\right|.\eqno(12)$$
Upon employing Eqs. (1), (11), and (12), we find
$$V_{\vk,\vk'} = - 5V_{2}\ {1\over 2}[3({\hat k}\cdot{\hat k'})^{2}-1]
N_{\vk}^{2}N_{\vk'}^{2}.\eqno(13)$$
The normalizing factor $N_{\vk}^{2}$ is given
$$N_{\vk}^{2} = {1\over 1 + |W_{\vk}|^{2}}, \eqno(14)$$
where $|W_{\vk}|^{2}$ is the relative probability contained in the
virtual spherical waves surrounding the impurities (compared to the
plane wave part).
As in s-wave superconductor case, in computing $|W_{\vk}|^{2}$, 
we cut off the radial integral at $R = 3.5\xi_{o}/2$, because
the pair-correlation amplitude falls exponentially as 
$exp(-r/3.5\xi_{o})$ near $T_{c}$.$^{23}$
The average correlation length $\xi_{o}$ is defined by 
$$\xi_{o} = 0.18 {v_{F}\over T_{c}}.\eqno(15)$$
With Eqs. (13), (14) and (15), one readily finds  
$$N_{\vk}^{2} = {1\over 1 + {3.5\xi_{o}\over 4\ell}},\eqno(16)$$
where $\ell$ is the mean free path. 

Therefore the pairing interaction is reduced: 
$$V_{\vk,\vk'} = - 5V_{2}\ {1\over 2}[3({\hat k}\cdot{\hat k'})^{2}-1]
 [1 + {3.5\xi_{o}\over 4\ell}]^{-2}.\eqno(17)$$
Notice that in dilute limit the reduction is proportional
to the ratio of the average correlation length to the mean free
path, $\xi_{o}/\ell$ and the pairing interaction decreases
linearly with the impurity concentration.
The $T_{c}$ equation is now,
$$T_{c} = 1.13\esn_{c}e^{-1/ N_{o}V_{2}[1+{3.5\xi_{o}\over 4\ell}]^{-2}}. 
\eqno(18)$$
Figure 1 shows $T_{c}$ versus $1/\tau$ for $T_{co}=40K\ \rm{and}\ 80K$ 
respectively. $T_{co}$ denotes the transition temperature without impurities.
We used $\epsilon_{c}=500K$. For a metal with $v_{F}=2\times 10^{7} cm/sec$,
the superconductivity is completely suppressed when the mean free paths are
about $1000 \AA$ and $350\AA$ for $T_{co}=40K\ \rm {and}\ 80K$, respectively.  
As in magnetic impurity effect on s-wave superconductors,$^{14}$ we find a 
sudden drop of $T_{c}$ near the critical impurity concentration. 
This effect was much weakened by potential scattering in s-wave case. 
In this case, this effect may be real. Note that Porto and Parpia$^{25}$ found a
sudden drop of $T_{c}$ in p-wave superfluid $He^{3}$ caused by aerogel. 
On the other hand, the inelastic scattering may decrease the effect in
high $T_{c}$ superconductors.

The change in $T_{c}$ relative to $T_{co}$ can be
calculated to first order in the impurity concentration. 
From Eqs. (15) - (18)
$$\Delta T_{c} \approx -{0.32\over \lambda\tau},\eqno(19)$$
where $\tau$ is the collision time due to impurities. 
The factor $1/\lambda$ shows that the initial slope (versus $1/\tau$)
depends on the superconductor and, consequently, is not a universal
constant. Weak superconductors lose their $T_{c}$ more rapidly
than strong ones. 

\vskip 1pc
\centerline{\bf 4. Magnetic Impurity Effect}
\vskip 1pc
The magnetic impurities give rise to both potential and magnetic 
scattering. For the potential scattering, the result of the
previous section is applicable.
In this section, we consider the effect of magnetic scattering.
The magnetic interaction between a conduction electron at $\bf r$ and a magnetic 
solute (having spin $\bf S$), located  at ${\bf R}_{i}$, is
$$H_{m} = J{\bs}\cdot{\bS}_{i}v_{o}\delta({\br}-{\bR}_{i}),\eqno(20)$$
where $\bs={1\over 2}\bf \sigma$ and $v_{o}$ is the atomic volume.
The three components of $\sigma$ are the Pauli matrices.
As in Sec. 3, we consider the magnetic impurity effect up to
the second order of $J$. 
Then,  the scattered state which carries the 
label, $\vk\alpha$, is 
$$\Psi_{\vk\alpha} = {\cal N}_{\vk}\Omega^{-{1 \over 2}} [ e^{i\vk\cdot \vr}\alpha
+ \sum_{\vq}e^{i(\vk + \vq)\cdot \vr}(W_{\vk\vq}\beta + W_{\vk\vq}^{'}
\alpha)], \eqno(21)$$
where,  
$$W_{\vk\vq} = {{1 \over 2}J\overline{S}v_{o}\Omega^{-1} \over 
\epsilon_{\vk} - \epsilon_{\vk+\vq}}\sum_{j}sin \chi_{j} e^{i\phi_{j}
-i\vq\cdot {\bR}_{j}},\eqno(22)$$
and,
$$W_{\vk\vq}' = {{1 \over 2}J\overline{S}v_{o}\Omega^{-1} \over
\epsilon_{\vk} - \epsilon_{\vk + \vq}}\sum_{j} cos \chi_{j} e^{-i\vq\cdot
{\bR}_{j}}. \eqno(23)$$
$\chi_{j}$ and $\phi_{j}$ are the polar and azimuthal angles of the spin
$\vS_{j}$ at ${\bR}_{j}$, and $\overline{S} = \sqrt{S(S+1)}$.
The degenerate partner of Eq. (21)
is:
$$\Psi_{-\vk\beta} = {\cal N}_{\vk}\Omega^{-{1 \over 2}}[e^{-i\vk\cdot\vr}\beta
+ \sum_{\vq}e^{-i(\vk + \vq)\cdot \vr} (W_{\vk\vq}^{*}\alpha - W_{\vk\vq}
^{'*}\beta)]. \eqno(24)$$
In this case, we pair $\Psi_{{\vec k}\alpha}$ and $\Psi_{-{\vec k}\beta}$.

Accordingly, the matrix element $V_{\vk, \vk'}$  is given
$$V_{\vk,\vk'} = <{\cal D}_{\vk'}(\vr_{1},\vr_{2})\big|V(|\vr_{1}-\vr_{2}|)
\big| {\cal D}_{\vk}(\vr_{1},\vr_{2})>,\eqno(25)$$
where
$${\cal D}_{\vk}(\vr_{1},\vr_{2}) = 
{1\over \sqrt{2}} \left|\matrix{\Psi_{\vk\alpha}({\br}_{1})&\Psi_{\vk\alpha}
({\br}_{2})\cr
\Psi_{-\vk\beta}({\br}_{1})&\Psi_{-\vk\beta}({\br}_{2})\cr}\right|.\eqno(26)$$
Using Eqs. (1) and (26), we find
$$V_{\vk,\vk'} = - 5V_{2}\ {1\over 2}[3({\hat k}\cdot{\hat k'})^{2}-1]
{\cal N}_{\vk}^{2}{\cal N}_{\vk'}^{2}.\eqno(27)$$
Note that Eq. (27) is the same form as Eq. (13) in Sec. II.
As a result, the $T_{c}$ reduction caused by the magnetic impurities 
can be calculated by the same formulas, Eqs. (17) - (19). The only difference 
is using $\ell_{s}$ and $\tau_{s}$ instead of $\ell$ and $\tau$. $\ell_{s}$ 
is the mean free path for the spin 
disorder scattering and $\tau_{s}$ is the spin-disorder scattering time. 
However, because the cross section for the potential scattering of the magnetic
impurities is generally much larger than that for exchange scattering,
we only need to consider the potential scattering due to
magnetic impurities as far as $T_{c}$ change is concerned.

\vskip 1pc
\centerline{\bf 5. Discussions}
\vskip 1pc

This study has been done in the framework of the BCS theory.
But the same result can be obtained from the Gor'kov's Green's function theory
if we impose a pairing constraint on the self-consistency equation.
Near $T_{c}$, the d-wave self-consistency equation with degenerate pairing
constraint is given by$^{26}$
$$\Delta^{*}({\bf r},{\bf l})=V({\bf r}-{\bf l})T\sum_{\omega}\int\int
d{\bf r}'d{\bf l}'\{G_{N}({\bf r}',{\bf r},-\omega)
G_{N}({\bf l}',{\bf l},\omega)\}^{P}
\Delta^{*}({\bf r}',{\bf l}').\eqno(28)$$
The superscript P denotes the pairing constraint and $G_{N}$
is the normal state Green's function in the presence of ordinary impurities.
Using the relation between $\Delta_{n}$ and $\Delta^{*}({\bf r},{\bf l})$,
$$\Delta_{n}=\int\int
d{\bf r}d{\bf l}\Delta^{*}({\bf r},{\bf l})\psi_{\bar n}({\bf l})\psi_{n}({\bf r}),\eqno(29)$$
it can be shown that Eq. (28) is the nothing but another form of the BCS 
gap equation.
We can also show that the physical constraint of the Anomalous Green's 
function $F^{\dagger}({\bf r}, {\bf l}, \omega)$, (i.e.),
$$\overline{F^{\dagger}({\bf r}, {\bf l}, \omega)}^{imp}=
\overline{F^{\dagger}({\bf r}-{\bf l}, \omega)}^{imp},\eqno(30)$$
gives rise to the degenerate pairing constraint. 
The superscript $\bar{\ }\bar{\ }^{imp}$ means an average over the 
impurity positions.

In high $T_{c}$ superconductors, the impurity doping 
and ion-beam induced damage$^{27}$ suppress strongly $T_{c}$. 
But the $T_{c}$ reduction 
is not fast enough to be explained by this study. 
The experimental data show that $T_{c}$ reduction is closely
related with the proximity to a metal-insulator transition caused
by the impurity doping and the ion-beam-induced damage.$^{20,21,27}$
We believe that the impurity response of the strongly correlated systems 
like high $T_{c}$ cuprates may be understood only when we consider 
the impurity scattering and the strong correlation on equal footing.
Because both non-magnetic and magnetic impurities suppress $T_{c}$ almost
equally, anisotropic pairing may be plausible.

\vskip 1pc
\centerline{\bf 6. Conclusion}
\vskip 1pc
In conclusion, we considered the effects of the (non-magnetic and magnetic)
impurities on $T_{c}$ in a d-wave superconductor.
The pairing interaction decreases linearly with the impurity concentration.
The initial slope of the $T_{c}$ suppression depends on the superconductor
and therefore is not a universal constant.
Because the cross section of the  potential scattering is much larger than
that of the exchange scattering, $T_{c}$ suppression
is determined entirely by the potential scattering of the magnetic impurities.  
We also discussed the implications of this study for the impurity doping
effect in high $T_{c}$ superconductors.

\vskip 1pc
\centerline{\bf Acknowledgments}
\vskip 1pc
This work has been supported by the Korea Research Foundation through 
Domestic Postdoctoral program. 
YJK acknowledges the supports by the Brain pool project of KOSEF and 
the MOST.
\vfill\eject
\centerline      {\bf REFERENCES} 
\vskip 1pt\hb
1. B. G. Levi, Phys. Today, May, 17 (1993), see also references therein. \hb 
2. D. J. Van Harlingen, Rev. Mod. Phys. {\bf 67}, 515 (1995).\hb
3. D. J. Scalapino, Phys. Rep. {\bf 250}, 329 (1995).\hb
4. P. J. Hirschfeld and N. Goldenfeld, Phys. Rev. B{\bf 48}, 4219 (1993).\hb
5. R. J. Radtke and K. Levin, H.-B. Schuttler, and M. R. Norman, Phys. Rev.
B{\bf 48}, 

653 (1993).\hb 
6. P. A. Lee, Phys. Rev. Lett. {\bf 71}, 1887 (1993).\hb 
7. T. Hotta, J. Phys. Soc. Jpn. {\bf 62}, 274 (1993).\hb
8. Y. Sun and K. Maki, Phys. Rev. B {\bf 51}, 6059 (1995).\hb
9. S. M. Quinlan, P. J. Hirschfeld, and D. J. Scalapino, Phys. Rev. B {\bf 53},
    8575 (1996).
10. A. A. Nersesyan, A. M. Tsvelik, and F. Wenger, Phys. Rev. Lett. {\bf 72},
    2628 (1994).\hb
11. K. Ziegler, M. H. Hettler, and P. J. Hirschfeld, Phy. Rev. Lett. {\bf 77},
    3013 (1996).\hb
12. S. Haas, A. V. Balatsky, M. Sigrist, and T. M. Rice, unpublished.\hb
13. A. A. Abrikosov and L. P. Gor'kov, Sov. Phys. JETP {\bf 12}, 1243 (1961).\hb
14. Yong-Jihn Kim and A. W. Overhauser, Phys. Rev. B{\bf 49}, 15779 (1994).\hb
15. M. F. Merriam, S. H. Liu, and D. P. Seraphim, Phys. Rev. {\bf 136}, A17 (1964).\hb 
16. W. Bauriedl and G. Heim, Z. Phys. B{\bf 26}, 29 (1977)\hb
17. A. Hofmann, W. Bauriedl, and P. Ziemann, Z. Phys. B{\bf 46}, 117 (1982).\hb
18. Yong-Jihn Kim, Mod. Phys. Lett. B {\bf 10}, 555 (1996).\hb
19. L. S. Borkowski and P. J. Hirschfeld, Phys. Rev. B {\bf 49}, 15404 
    (1994).\hb
20. G. Xiao, M. Z. Cieplak, J. Q. Xiao, and C. L. Chien, Phys. Rev. B {\bf 42},
    8752 (1990).\hb
21. V. P. S. Awanda, S. K. Agarwal, M. P. Das, and A. V. Narlikar, J. Phys.:
    Condens. 

Matter {\bf 4}, 4971 (1992).\hb
22. Yong-Jihn Kim and K. J. Chang, to appear in J. Kor. Phys. Soc. 
    Supplement (1997). \hb  
23. P. W. Anderson and P. Morel, Phys. Rev. {\bf 123}, 1911 (1961).\hb
24. P. W. Anderson, J. Phys. Chem. Solids {\bf 11}, 26 (1959).\hb
25. J. V. Porto and J. M.Parpia, Phys. Rev. Lett. {\bf 74}, 4667 (1995).\hb
26. W. Xu, Y. Ren, and C. S. Ting, Phys. Rev. B {\bf 53}, 12481 (1996).\hb
27. Y. Li, G. Xiong, and Z. Gan, Physica C {\bf 199}, 269 (1992).\hb
\vfill\eject
\centerline      {\bf Figure Captions} 
\vskip 1pt\hb
{\bf Fig. 1} Variation of $T_{c}$ with impurity concentration (measured 
in terms of the scattering rate, $1\over \tau$) for $T_{co}=40K$ and
$80K$, respectively. The cutoff energy $\epsilon_{c}$ is $500K$.
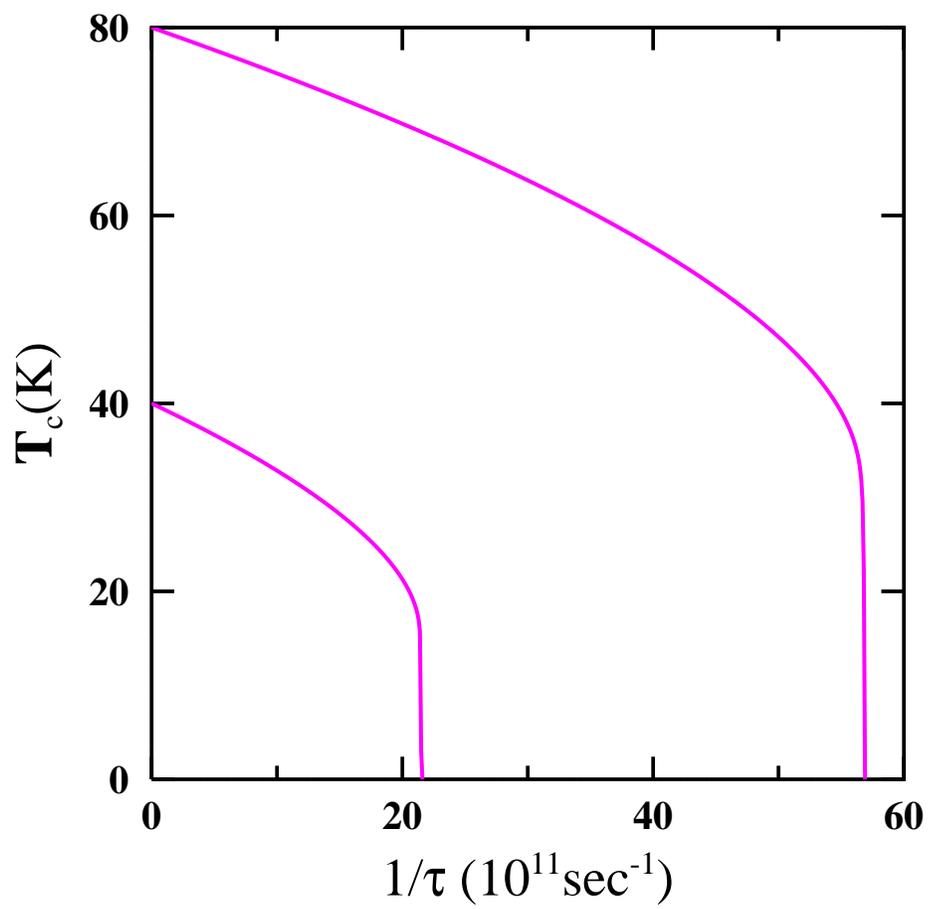
\end